\documentclass[letterpaper, 10 pt, conference]{ieeeconf}  

\IEEEoverridecommandlockouts                              
\usepackage[utf8]{inputenc}
\usepackage[T1]{fontenc}
\usepackage{amsmath}
\usepackage{comment}
\usepackage{verbatim}  
\usepackage{subcaption}
\usepackage{graphicx}
\usepackage{amsmath,amssymb,amsfonts}
\usepackage{mathtools}
\usepackage[super]{nth}
\usepackage{pdfpages}
\usepackage{algorithmicx}
\usepackage[ruled,vlined,linesnumbered]{algorithm2e}

\usepackage{subfig}
\usepackage{float} 

\title{\LARGE \bf
Transforming Next-generation Network Planning assisted by Data Acquisition of Top Three Spanish MNOs
}

\author{M. Umar Khan$^{1}$ 
\thanks{*This work was not supported by any organization}
\thanks{$^{1}$M. Umar Khan is with Faculty of Computer Engineering, COMSATS University Islamabad, Pakistan
        {\tt\small umar\_khan@comsats.edu.pk, www.drmukhan.com}}%
\thanks{}%
}

\begin{document}

\maketitle
\thispagestyle{empty}
\pagestyle{empty}

\begin{abstract}

In this paper, we address the necessity of data related to mobile traffic of the legacy infrastructure to extract useful information and perform network dimensioning for 5G. These data can help us achieve a more efficient network planning design, especially in terms of topology and cost. To that end, a real open database of top three Spanish mobile network operators (MNOs) is used to estimate the traffic and to identify the area of highest user density for the deployment of new services. We propose the data acquisition procedure described to clean the database, to extract meaningful traffic information and to visualize traffic density patterns for new gNB deployments. We present the state of the art in Network Data. We describe the considered network database in detail. The Network Data Acquisition entity along with the proposed procedure is explained. The corresponding results are discussed, following the conclusions. 

\end{abstract}

\section{INTRODUCTION}

Cellular network data has received attention even in the earlier studies to analyze the spatial and temporal variations of the subscribers \cite{data2}, to identify network traffic patterns \cite{data1} and to enhance the QoE for the end user \cite{data3}. Research on network data has a multi-disciplinary nature covering traffic management aspects, service provisioning and city dynamics in relation to signal processing, mobile communications and data mining parameters \cite{newdata1}. The availability of network data can be advantageous to achieve enhanced performance with a profound impact on the network design, cost and topology \cite{newdata2}. However, an in-depth knowledge is required to understand the network data sources, accessibility, features and applications to extract related information for a particular use case \cite{newdata3}. Moreover, these studies are conducted on the data obtained from different sources such as mobile phones, deployed sensors, cloud radio access network (C-RAN) and data collected by the corresponding MNO itself. The data owned by MNOs or obtained from C-RAN are private because they contain sensitive information corresponding to the operator and may not be publicly available. However, thanks to the existence of openly accessible crowdsourced databases, researchers have access to data that make it possible to analyze and compare the performance of different networks and operators without resourcing to data owned by MNOs. For instance, in \cite{fmand} the authors utilize a real-world crowdsourced database to analyze legacy LTE deployments and the traffic they serve. The proposed methodology is designed to identify the existing deployments with the objective of handling present-day and future traffic demands and the possibilities to improve the network capacity. The results show that the deployed LTE cells are over-provisioned compared to the current traffic needs, even though a considerable capacity enhancement will be needed to serve the future traffic load. The work of \cite{zoraida_data} proposes a methodology based on a crowdsourced database to infer the status of the existing infrastructure, and to analyze the network density and spectrum resources of multiple MNOs. The results indicate that in dense urban areas, the LTE cells deployment is capacity-oriented to deal with higher traffic demands. On the other hand, the deployed resources for the suburban areas are coverage-oriented by deploying lower frequency spectrum in order to minimize the deployment cost. The findings obtained from the proposed methodology are discussed with reference to the design of new 5G deployments.

In this context, crowdsourcing refers to obtaining input from the community or specific working groups. More specifically, data crowdsourcing involves data acquisition, management and analysis to implement an appropriate and meaningful solution of a particular problem \cite{crowdsourced_main,crowdsourced_main1}. The use of open, crowdsourced data has broaden the scope of the research community ranging from planning smart cities to business development or analytics \cite{cr1}. In wireless communications, numerous studies have already presented significant findings by leveraging crowdsourced databases. For instance, the authors of \cite{data_sec, data_sec1} propose a data collection and analysis model to measure the security level against different threats faced by the LTE network. In \cite{crd1}, the authors prefer a crowdsourced database for assessment and continuous monitoring of electromagnetic field (EMF) exposure in LTE cells compared to conventional methods. They show that a suitable estimate of the EMF strength can be provided from the measurements of signal strength obtained from crowdsourcing. A study on a real world crowdsourced database from a mobile application identifies multiple aspects of traffic \cite{maldarino1} such as the contribution of vehicular traffic to the global traffic and performance of different architectures by getting useful insights from traffic data. The authors in \cite{cr2} propose a model that analyzes the LTE throughput on the down-link with relation to other network variables such as time, signal strength, signal quality and MNO. Through data analytics they conclude that signal quality compared to signal strength is the appropriate variable to predict the LTE user throughput. 

At the same time, the benefit of using crowdsourced data is that a range of parameters and indicators are available to deduce meaningful information. For instance evolved Node-B (eNB) density, reference signal received power (RSRP), reference signal received quality (RSRQ) and available bandwidth are scrutinized to observe how the allocated spectrum is used in LTE networks in different geographical areas \cite{zoraida_data_cr4}. In \cite{cr3}, the authors study the legacy infrastructure to infer if the previous LTE deployment strategies of the MNOs were driven by coverage or capacity in the target area. The analysis reveal that most of the MNOs' strategies were aimed at providing the coverage in the target area rather than the capacity enhancements. 

In the light of the previous discussion, we observe that there are several aspects regarding network data that can be taken into consideration to render impactful insights in the network planning phase. In this regard, we have noticed that the data acquired from an open database can be utilized to identify the area of highest traffic density, which is a relevant input to network dimensioning. This also entails an additional benefit in terms of network planning cost. In this study, network data from the OpenCellID database is used to obtain information about the state of the legacy network of MNOs related to traffic density. The OpenCellID database is chosen because millions of samples from existing cellular networks are publicly accessible for the research community and these samples relate LTE identifiers with traffic density. Therefore, in this paper we focus on utilizing OpenCellID, whose details are provided in \S \ref{sec:OpenCellID}.

\subsection{Contributions}
This paper presents a data acquisition approach which enables the demarcation of the target deployment area for new 5G services. In brief, this paper brings the following contributions: 

\begin{enumerate}
	\item We propose the utilization of real network data to estimate network traffic and the target deployment area where the subscribers concentration is maximum, reflecting the most congested area. 
	\item We develop and propose a data acquisition procedure. The proposed procedure utilizes the standard LTE identifiers to determine the highest traffic area.
	
	\item We have evaluated our data analysis procedure on data from the top three Spanish MNOs.  
\end{enumerate} 

The following sections describe the complete process for network data acquisition. We first describe the crowdsourced database in detail and then discuss the procedure. 
\section{The OpenCellID Database}\label{sec:OpenCellID}

The OpenCellID is an open-access database/platform to which mobile devices send data related to their connection (e.g. position and different identifiers) through a crowdsourcing application \cite{opencellid}. This application records the global positioning system (GPS) coordinates (latitude, longitude) of both device and base station (BS), the radio access technology (RAT), mobile country code (MCC), mobile network code (MNC), tracking area code (TAC), cell tower ID (CID), approximate area in meters where the cell possibly exists (range), and the number of samples recorded. When a device establishes a connection with the base station, it scans all the information of the above variables and each scan, denoted as \textit{sample}, is recorded by the device and submitted to the database. The database also provides information about the BS location, whether it is obtained from the samples or provided to the database by the MNO itself.

\subsection{The Database Matrix}
The OpenCellID database contains raw data that need to be processed for its correct use. We consider raw data belonging to Spain in order to understand network infrastructure that would be useful for new technology deployments according to the 5G national plan \cite{Spanishplan}. First we form the network database matrix $\mathbf{D}$ by extracting information corresponding to MCC=214 for Spain from the raw database. In matrix $\mathbf{D}$, $E$ is the number of samples in millions, and each row $\mathbf{d}_e$ of $\mathbf{D}$ represents a  data point $\mathbf{d}_e = (r_e, m_e, t_e, n_e, l_e, s_e)$. Hence, $\mathbf{D}:=\{ \mathbf{d_e}\}_{e=1}^E ~\forall ~\mathbf{d_e} \in \mathbb{R}^v$, where $v=6$ represents the number of data variables   

\begin{equation*}
	\mathbf{D} = 
	\begin{pmatrix}
		r_{1} & m_{1} & t_{1} & n_{1}  & l_{1}  & s_{1}  \\
		r_{2} & m_{2} & t_{2} & n_{2}  & l_{2}  & s_{2}  \\
		\vdots&\vdots &\vdots &\vdots  &\vdots    &\vdots  \\
		r_{e} & m_{e} & t_{e} & n_{e}  & l_{e}  & s_{e}  \\
		\vdots&\vdots &\vdots &\vdots  &\vdots    &\vdots  \\
		r_{E} & m_{E} & t_{E} & n_{E}  & l_{E}  & s_{E}	
	\end{pmatrix},
\end{equation*}

and the components of the data point $\mathbf{d_e}$ are:

\begin{itemize}
	\item $r_e$: radio access technology (RAT) of row  $e$.
	\item ${m}_e$: mobile network code (MNC) of row $e$.
	\item ${t}_e$: tracking area code (TAC) of row $e$.
	\item ${n}_e$: cell ID (CID) of row $e$.
	\item ${l}_e$: bi-dimensional location of the cell with latitude and longitude $l_e=(l^1_e,l^2_e)$. 
	\item ${s}_e$: number of samples of row $e$. 
\end{itemize}

The dataset contains the coordinates of the existing GSM, UMTS and LTE cells in Spain. In this study, we consider the legacy LTE cells as they are the source to estimate current data utilization in the cellular networks, given that GSM and UMTS make a very little contribution to the present-day traffic. The dataset contains millions of CIDs which are sufficient for network analysis to deduce the infrastructure and traffic information. However, the eNB-IDs are not provided by the database to preserve the privacy of MNOs. On the other hand, the accessibility to eNB-IDs can ease the process of data acquisition and traffic information can be made promptly available for analysis. In the absence of eNB-IDs, there exists the need for a data acquisition procedure that extracts the traffic information from uniquely identified cells.

\section{The Data Acquisition Procedure}\label{sec:Data_acquisition}

\begin{figure}[h]
	\centering  
	\begin{center}           
		\includegraphics[width=.450\textwidth]{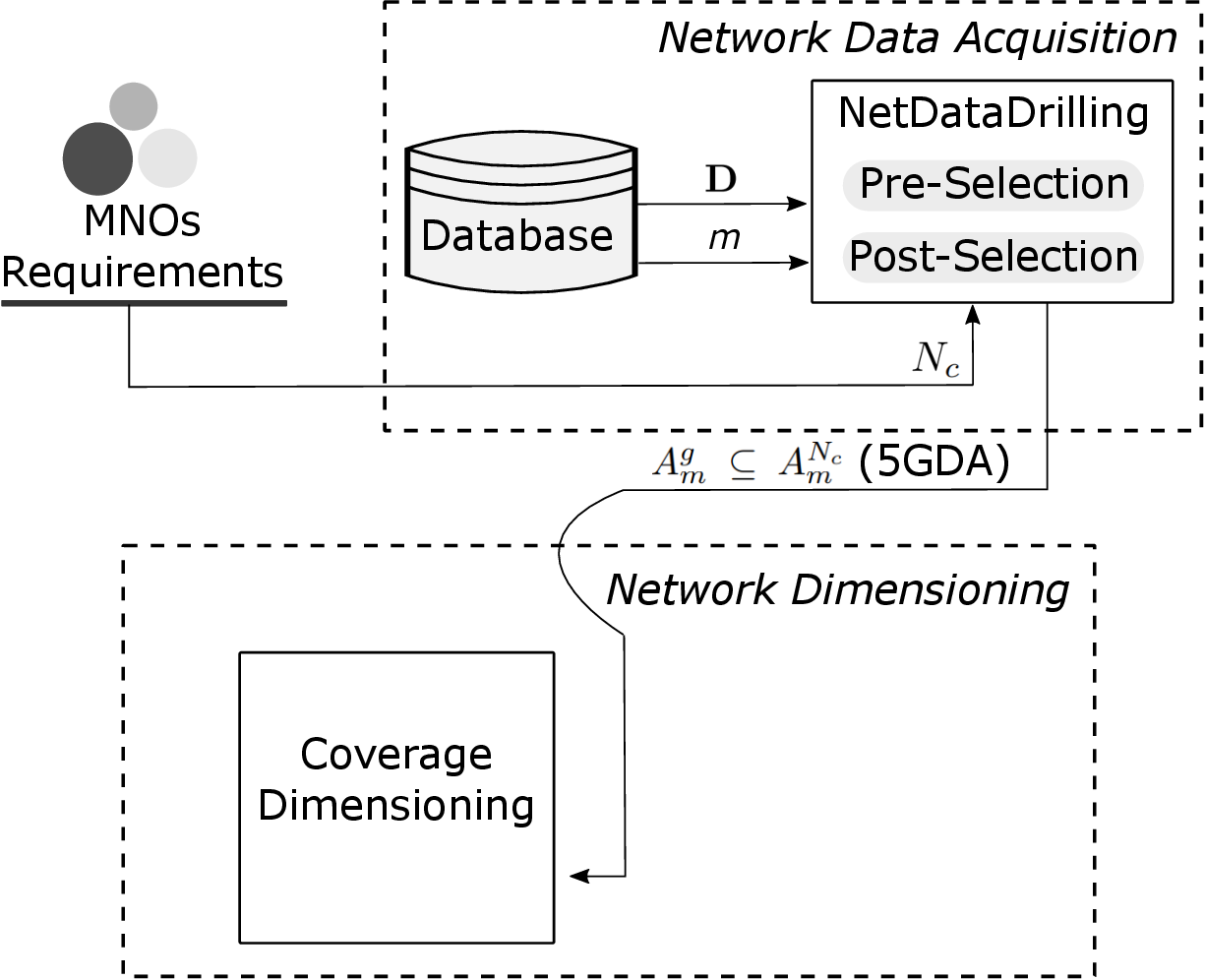}
	\end{center} 
	\caption{Network Data Acquisition entity linked with Network Dimensioning to provide 5GDA ($A_m^{g}$) as an input to Coverage Dimensioning}\label{fig:net_data_acq}
\end{figure}

In the previous section, we introduced how we form the matrix of row data $\mathbf{D}$ from the OpenCellID database. However this information must be processed to serve a specific goal. One possible goal is identifying the 5G deployment area (5GDA) with maximum density of subscribers, that would represent the region with the highest traffic with high probability. This section introduces the procedure Network Data Acquisition of Fig.\ref{fig:net_data_acq}. The core of this procedure is determining the tracking area that contains the largest number of samples and this tracking area is identified with the highest traffic TAC. On the other hand, we also need to guarantee the reliability of the dataset, for instance to avoid duplicates in the selected base stations. To that end, the \textit{NetDataDrilling} (NDD) procedure in \S \ref{sec:Net_Data_Drilling} is designed. This procedure consists of two phases, Pre-Selection and Post-Selection described in \ref{sec-Pre-Selection} and \ref{sec-Post-Selection}, respectively.



\subsection{Highest Traffic TAC Area}
The highest traffic TAC (HTTAC) area consists of those cells containing the highest number of samples $s_e$ provided by the network database $\mathbf{D}$. In $\mathbf{D}$, each MNO $m$ have a set of TACs $\mathcal{T}_m = \left\{t_1,\ldots,t_j,\ldots,t_J\right\}$, where $t_j$ represents an individual TAC; a set of cells $\mathcal{N}_m = \left\{n_1,\ldots,n_k,\ldots,n_K \right\}$, where $n_k$ represents the CID and $N_m=|\mathcal{N}_m|$ is the total number of cells for MNO $m$; and $s_m^{j,k}$ stands for the number of samples from cell $k$ inside the $j$th TAC.\footnote{Note that $s_m^{j,k}=0$ if the cell is not part of the tracking area.} We assume that the highest number of samples per cell represents the highest traffic density in that area. Therefore, the ID of the highest traffic TAC $H_m$ or HTTAC is obtained from the maximum aggregated number of samples of all existing cells inside the tracking area TAC as:

\begin{equation}
	\begin{split}
		H_m & = \arg_j \max H_m^j, \\
		H_m^j & = \sum_{k=1}^K s_m^{j,k}, \forall j=1,\ldots,J.
	\end{split}
\end{equation}

\subsection{5G Deployment Area (5GDA)}
The deployment area refers to the demarcation of a geographical region for the installation of new BSs to offer new services. As 5G networks must offer new services with a lower cost per bit compared to LTE networks, a larger number of subscriptions increases network utilization and, at the same time, help MNOs to reduce the cost per bit. Moreover, each MNO may have its own roll-out strategy concerning the deployment cost of the 5G infrastructure. Therefore, it is pertinent to identify the highest traffic area. We proceed by selecting $N_c$ legacy cells inside $H_m$ such that $N_c \subseteq \mathcal{N}_m$, being $N_c$ those cells supporting the highest traffic density. These cells provide an estimate of the highest traffic areas where most of the subscribers are localized so that an MNO may start deploying new 5G services in order to minimize the cost. The geographical area covering the $N_c$ cells is represented by $A_m^{N_c}$. Besides, MNOs may have financial constraints for the deployment of 5G services in that area which is addressed through the value of $N_c$ cells. If a larger value of $N_c$ is selected, the area $A_m^{N_c}$ will be larger and the MNO has to deploy its services in a wider area, leading to a higher cost and the un-fulfillment of the financial budget. Therefore, the value of $N_c$ translates into the financial and coverage area limitations of the MNO. Moreover, we define $A_m^g$ as the area of high traffic concentration which is determined through data visualization where the gNBs will be deployed for 5G services such that $A_m^g \subseteq A_m^{N_c}$.


\subsection{Pre-Selection} \label{sec-Pre-Selection}

The Pre-Selection phase is the first phase of the NDD procedure (see Fig.\ref{fig:net_data_acq}) and extracts information for a particular country (MCC) and its networks (MNC) from the global database $\mathbf{D}$. We then identify the top MNOs and check that the RATs provided in the database are correct by comparing with a external data source, as described below:

\begin{itemize}	
	\item First, we extract the subset of network data belonging to Spain (country code MCC=214). 
	\item We select the top three MNOs through their respective market shares in Spain \cite{statista}.    
	\item We confirm that the LTE RATs provided by the OpenCellID database have been actually deployed by the MNOs by contrasting this information with \cite{Halber}. 
\end{itemize} 

Note that data from any other country can be processed by simply selecting a different MCC. Similarly, the number of top MNOs can be customized.

\subsection{Post-Selection} \label{sec-Post-Selection}
The Post-selection phase is the core of the NDD procedure, ensuring the quality of our analysis, avoiding data redundancy, and revealing patterns to recognize the areas with higher density of subscribers. To ensure the quality of our analysis, we define the confidence benchmark samples (CBS) interval as the number of samples of legacy cells in the range $[a,b]$ so that the existence of the $N_c$ cells is guaranteed. Note that in the database there are cells whose recorded number of samples is small (e.g., lower than 100) and this situation may well represent cells that were installed and removed, reconfigured or merged in a subsequent stage. There are also a few cells that have a very large number of samples compared to the majority of cells (more than 2000), which are considered outliers. Therefore, we do not include those cells in our analysis by defining $a$ and $b$ as the lower and upper bound, respectively. 

Next, we combine two LTE identifiers to uniquely determine the cells, referred to as TAC and CID combination. Once the cells are uniquely identified, we obtain the region with highest traffic area. From the HTTAC area, we identify the cells where the MNOs select a feasible sub-area, the 5GDA. The 5GDA is provided as an input to the Network Dimensioning entity (see Fig.\ref{fig:net_data_acq}) for the Coverage Dimensioning. 

The Post-selection phase contemplates the following aspects:

\begin{itemize}	
	\item There exist multiple samples of a single cell in the database. A single CID may correspond to multiple recorded GPS coordinates. To avoid redundancy of multiple cells at the same true location, we define the combination of TAC and CID to take individual cells. 
	\item To estimate the actual traffic of each MNO, we assume that the areas (TAC) with the highest number of samples support the highest traffic density in that area, referred to as HTTAC.
	\item We then select the $N_c$ cells inside the HTTAC such that these cells have the highest number of samples, i.e. the number of connections established by UEs.
	\item Finally, the 5GDA sub-area $A_m^g  \subseteq  A_m^{N_c}$ is selected, representing the highest traffic density area. 
\end{itemize} 




\subsection{The NetDataDrilling Algorithm} \label{sec:Net_Data_Drilling}
The procedure \textit{NetDataDrilling} (NDD) for data selection and extraction is described in this subsection and its pseudocode is provided in Algorithm \ref{ALGO:net_data_drilling}. It requires the following inputs:
\begin{itemize}
	\item The network database $\mathbf{D}$.
	\item The country code represented by MCC $M$.
	\item The constant $N_c$ for the selection of legacy cells. 
	\item The top three MNOs represented with MNCs $m$. 
	\item The constants $a$ and $b$ as lower and upper bound of the CBS approach.
\end{itemize}

The outcome of the NDD procedure is $H_m$, the HTTAC ID for each MNO $m$, and the set of cells $\mathcal{N}_{c}$ with the $N_c$ CIDs.

In step 5, the procedure Preselection extracts the database corresponding to country code $M$ and the $m$ MNOs. In the procedure, performed for each MNO (\textit{for} loop in step 6), some auxiliary variables store intermediate results. $S_{aux}$ represents the summation of the samples of those cells within each TAC, then having a dimension equal to the number of tracking areas. $V_{aux}$ represents the summation of samples of each cell inside the HTTAC (with dimension equal to the number of cells forming the HTTAC) and the CIDs of these cells are stored in $W_{aux}$.

We first remove the data points $\mathbf{d}_e$ such that their number of samples is out of the CBS bounds, i.e. $a>s_e>b$ (steps 7-11). Next, we find the unique TAC and CIDs. Since there may exist multiple versions of the same cell (same CID) due to different recorded GPS coordinates, the combination of TAC and CID determines unique TACs and CIDs and eliminates redundancy. The process UNIQUE uses the pair (TAC, CID) to identify uniqueness and store the indices of TACs and CIDs in $\mathcal{T}_m$ and $\mathcal{N}_m$, respectively (step 12).

\begin{algorithm}[h!] \label{algo_NetDataDriling}
	\SetAlgoLined
	\text{\textbf{procedure}} {NetDataDrilling}$(\mathbf{D},M,N_c,m,a,b)$\\
	$S_{aux} \leftarrow \mathbf{0}$ \\		
	$V_{aux} \leftarrow \mathbf{0}$ \\
	$W_{aux} \leftarrow \mathbf{0}$ \\
	$\left[m,\mathbf{D} \right] \leftarrow \text{Preselection}(M,\mathbf{D}) $\\
	\For{\textbf{all} $m$}{
		\For{\textbf{all} $e$}{
			\If{($s_e < a \lor s_e > b$)}{
				$\mathbf{D} \leftarrow  \mathbf{D}\backslash \mathbf{d}_e$}
		} 
		$\left[\mathcal{T}_m, \mathcal{N}_m \right] \leftarrow \text{UNIQUE}(\mathbf{D}(t_e,n_e)) ~\forall e$ \\
		\For{\textbf{all} $t_e \in \mathcal{T}_m$\\
			$k\leftarrow j: t_e=\mathcal{T}_m(j)$}{
			\For{\textbf{all} $e$}{
				\If{$t_e=\mathbf{D}(t_e)$ and $\mathbf{D}(n_e) \in \mathcal{N}_m$}{
					{$S_{aux}(k) \leftarrow S_{aux}(k) + s_e$}
				}
			}			
		}
		
		$H_m=\max S_{aux}$\\
		\For{\textbf{all} $e:\mathbf{D}(t_{e})=H_m$}{
			Find index $k:\mathcal{N}_m(k)=\mathbf{D}(n_{e})$\\
			$V_{aux}(k) \leftarrow V_{aux}(k) + s_e$\\
			$W_{aux}(k) \leftarrow \mathcal{N}_m(k)$ \\
		}
		
		$\mathcal{N}_c\leftarrow \text{SORT} (V_{aux}, W_{aux})$\\
		$\mathcal{N}_c\leftarrow \{n_{c,1},\ldots,n_{c,k},\ldots,n_{c,N_c}\}$
		
	}
	
	\caption{NetDataDrilling} \label{ALGO:net_data_drilling}
\end{algorithm}

In steps 13-20, the total number of samples per TAC is stored in $S_{aux}$ (step 17) by summing the samples of those CIDs associated with each TAC. We begin the loop in step 14, keeping in $k$ the index of the TAC in $\mathcal{T}_m$ (given by $j$) to the TAC corresponding to the TAC ($t_e$) of that iteration $e$ of the loop. For every TAC ID in $\mathcal{T}_m$, we match their IDs with those in the database $\mathbf{D}(t_e)$ and check whether the cell of the analyzed row $e$, denoted by $\mathbf{D}(n_e)$, is part of the MNO's network (step 16). Here, $\mathbf{D}(t_{e})$ and $\mathbf{D}(n_{e})$ from $\mathbf{D}$ represent respectively the TAC ID (\nth{3} column) and the CID (\nth{4} column) in row $e$. In step 21, we find the tracking area with the highest traffic $H_{m}$ for MNO $m$, corresponding to the TAC with the maximum number of recorded samples.

In steps (22-26), we calculate the total number of samples of the cells residing inside the target area $H_{m}$. So, once $H_{m}$ is found in $\mathbf{D}$, the number of samples corresponding to the cells within $H_{m}$ are stored in $V_{aux}$ (step 24). The CIDs of the cells residing inside the target area $H_{m}$ are then stored in $W_{aux}$ (step 25). Finally, we need to identify the $N_c$ cells supporting the highest traffic. In step 27, the cells corresponding to $H_{m}$ are sorted according to the number of samples in decreasing order. Then, as shown in step 28, the selected $N_c$ cells are the first $N_c$ positions of the set $\mathcal{N}_c$, where $n_{c,k}$ denotes the CID of the cell with $k$-th highest traffic within $H_m$.

\subsubsection{Complexity of the NDD Algorithm}
The complexity of Algorithm \ref{ALGO:net_data_drilling} is based on the size of the database $\mathbf{D}$ represented by the number of samples $E$. To process $\mathbf{D}$, a finite number of iterations is required given by the \textit{for} loops. Therefore, the complexity of Algorithm \ref{ALGO:net_data_drilling} represented by the inner loops can be calculated as:
\begin{equation}\label{eq:compl1}
	\mathcal{O}(E) + \mathcal{O}(E \times |\mathcal{T}_m|) + \mathcal{O}(|\mathbf{D}(t_e)|).
\end{equation}	
Note that (\ref{eq:compl1}) provides the complexity required for each MNO $m$ and, in practice, this would be the total complexity as the samples from each MNO can be independently processed in parallel.

\section{Results}\label{sec:network_data_results}
\begin{figure}[h]
	\centering
	\includegraphics[width=.48\textwidth]{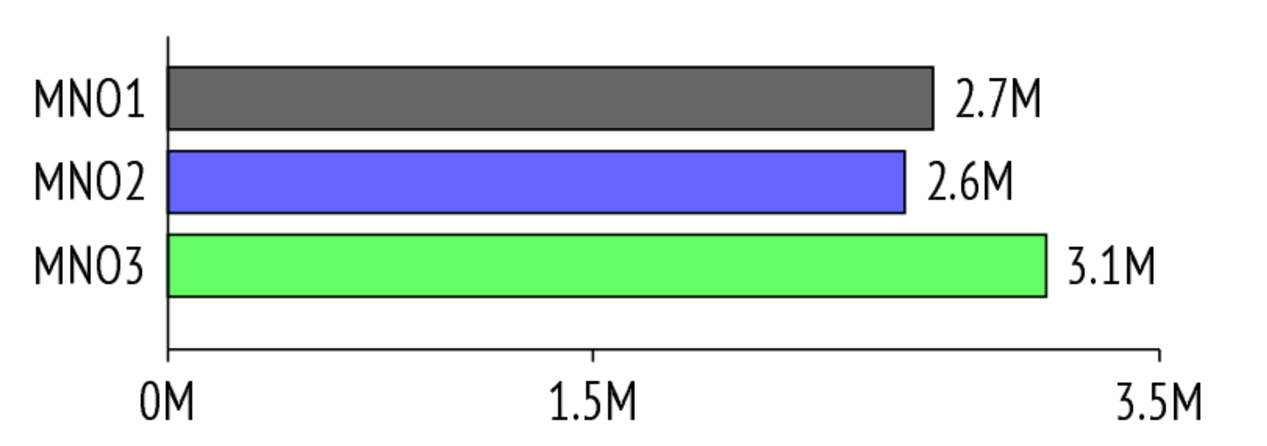}
	\caption{Accumulative samples (GSM + UMTS + LTE) for top three Spainish MNOs} \label{fig:MNOs_samples}
\end{figure}

\begin{figure*}[p] 
	\centering
	\begin{minipage}{0.48\textwidth}
		\centering
		\begin{subfigure}{\textwidth}
			\includegraphics[trim={0 0cm 0 .8cm}, clip, width=\linewidth]{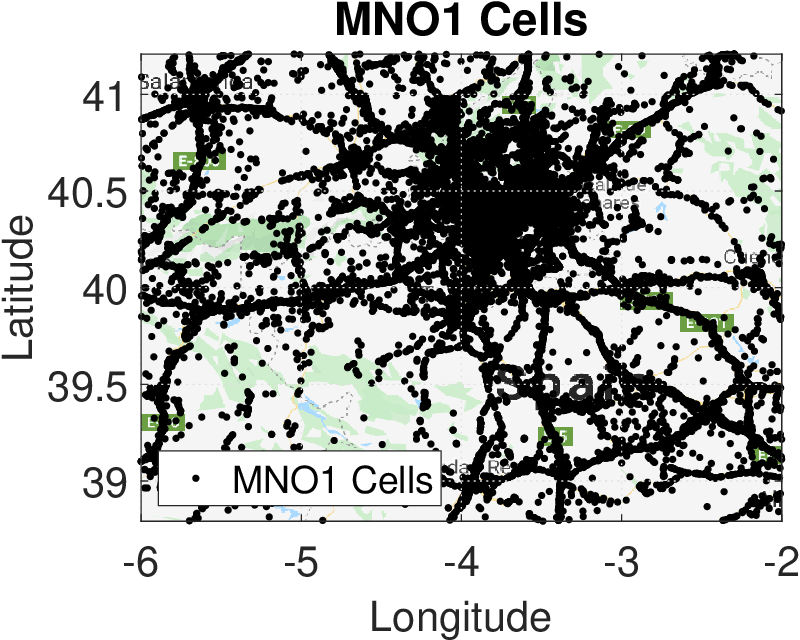}
			\caption{}
			\label{subfig:mnc1}
		\end{subfigure}
		
		\vspace{\baselineskip} 
		
		\begin{subfigure}{\textwidth}
			\includegraphics[trim={0 0cm 0 .8cm}, clip, width=\linewidth]{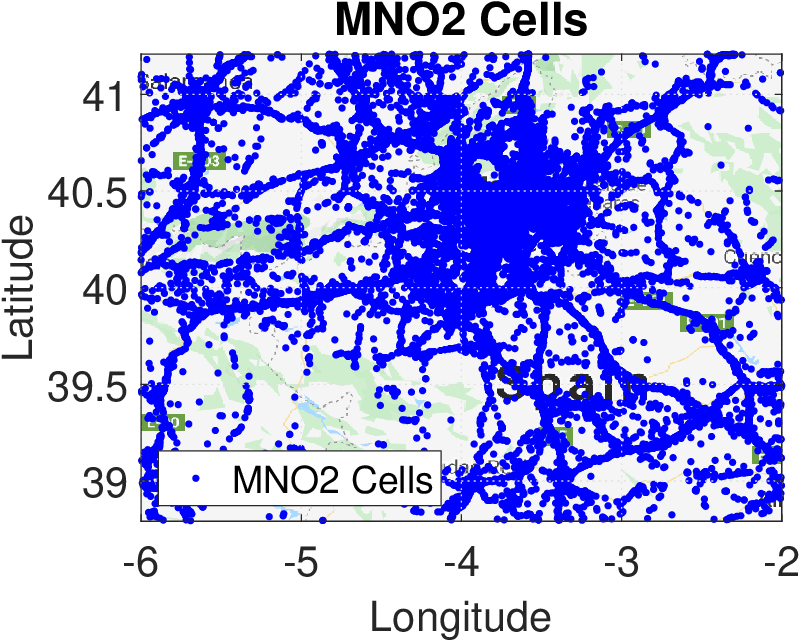}
			\caption{}
			\label{subfig:mnc2}
		\end{subfigure}
		
		\vspace{\baselineskip}
		
		\begin{subfigure}{\textwidth}
			\includegraphics[trim={0 0cm 0 .8cm}, clip, width=\linewidth]{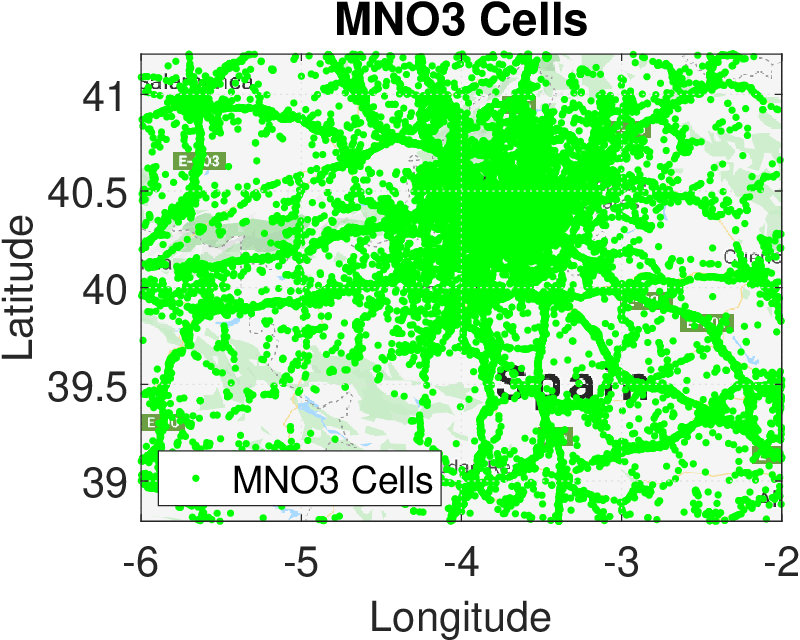}
			\caption{}
			\label{subfig:mnc3}
		\end{subfigure}
	\end{minipage}%
	\hfill
	\begin{minipage}{0.48\textwidth}
		\centering
		\begin{subfigure}{\textwidth}
			\includegraphics[trim={0 0cm 0 .9cm}, clip, width=\linewidth]{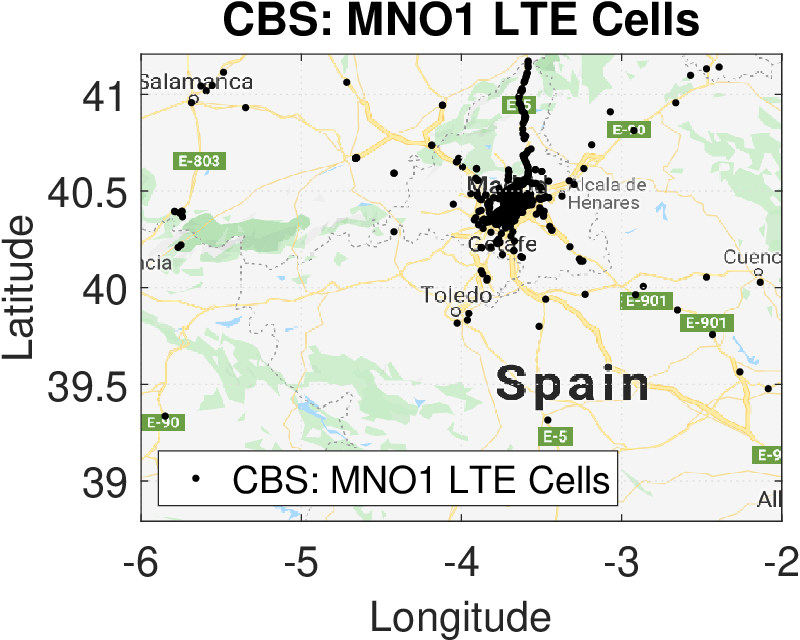}
			\caption{}
			\label{subfig:mnc1_lte}
		\end{subfigure}
		
		\vspace{\baselineskip}
		
		\begin{subfigure}{\textwidth}
			\includegraphics[trim={0 0cm 0 .9cm}, clip, width=\linewidth]{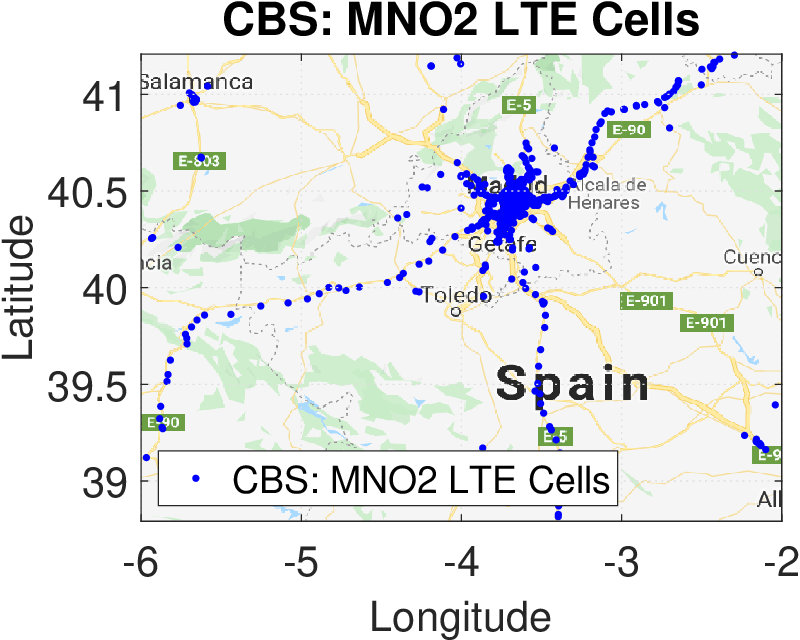}
			\caption{}
			\label{subfig:mnc2_lte}
		\end{subfigure}
		
		\vspace{\baselineskip}
		
		\begin{subfigure}{\textwidth}
			\includegraphics[trim={0 0cm 0 .9cm}, clip, width=\linewidth]{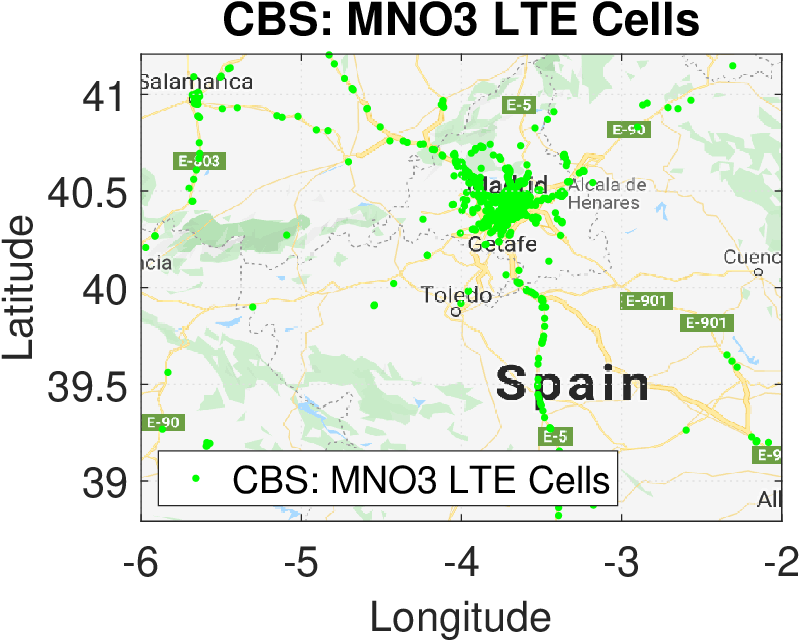}
			\caption{}
			\label{subfig:mnc3_lte}
		\end{subfigure}
	\end{minipage}
	
	\caption{Accumulative samples (GSM + UMTS + LTE) (a,c,e) and LTE samples as a result of CBS for three Spanish MNOs (b,d,f)} 
	\label{fig_measurements}
\end{figure*}

The first step is the creation of the database $\mathbf{D}$, used by the NDD algorithm, from OpenCellID data. The accumulative samples for Spain corresponding to GSM + UMTS + LTE recorded per MNO are 2.7, 2.6 and 3.1 millions (see Fig.\ref{fig:MNOs_samples}) for MNO1, MNO2 and MNO3, respectively, which are significant enough for our analysis. We plot the samples corresponding to three MNOs as shown on the map in Fig.\ref{fig_measurements}(a,c,e) by plotting the cell base stations.

The NDD procedure in Algorithm \ref{ALGO:net_data_drilling} imposes the CBS approach by setting $a=100$ and $b=1000$. The LTE cells fulfilling the CBS criterion are shown in Fig.\ref{fig_measurements}(b,d,f). 
These cells are in thousands (3938, 3404, 3665) referring to MNO1, MNO2 and MNO3, respectively. Unique TACs and CIDs are now identified. Checking of individual LTE cells is applied to the region plotted in Fig.\ref{fig_measurements}(b,d,f) with the combination of unique TAC ($\mathcal{T}_m$) and CIDs ($\mathcal{N}_m$) to obtain the number of unique LTE cells. The output of the combination shows that MNO1 has 3895, while MNO2 and MNO3 have 3296 and 3408 cells, with market shares $\kappa_1=23.55\%$, $\kappa_2=25.79\%$ and $\kappa_3=30.14\%$, respectively \cite{statista}.

\begin{figure*}[htbp]
	\centering 
	
	\begin{subfigure}{0.45\textwidth}
		\includegraphics[trim={0 0cm 0 .9cm},clip, width=\linewidth]{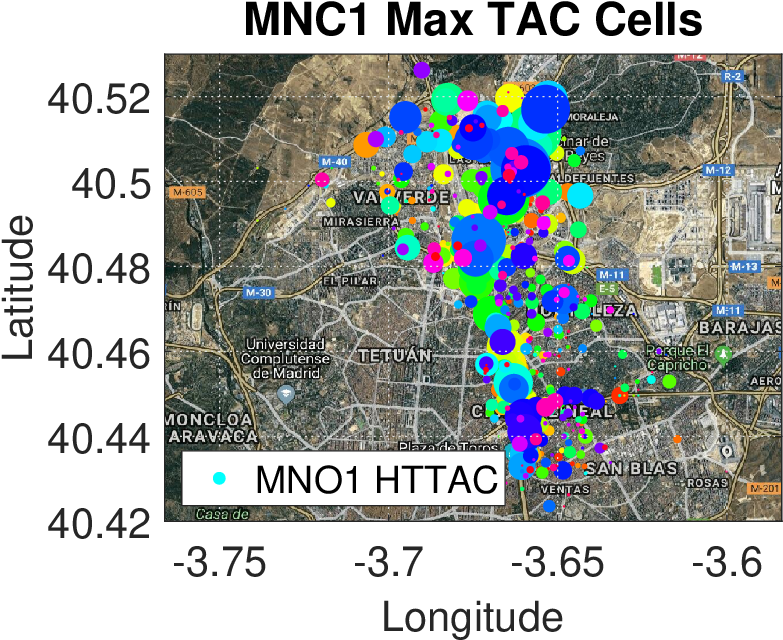}
		\caption{}
	\end{subfigure}
	\hfill
	\begin{subfigure}{0.45\textwidth}
		\includegraphics[trim={0 0cm 0 0.9cm},clip, width=\linewidth]{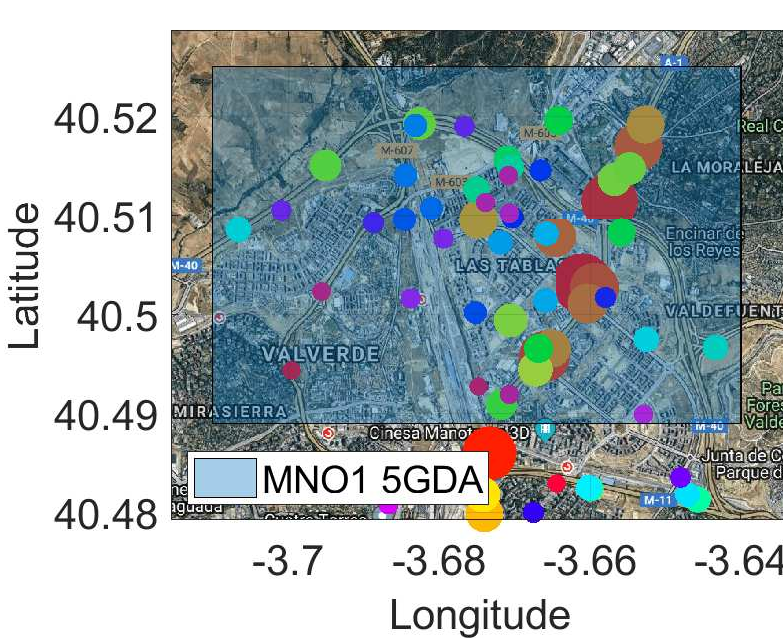}
		\caption{}
	\end{subfigure}
	
	\vspace*{-\baselineskip} 
	\begin{subfigure}{0.45\textwidth}
		\includegraphics[trim={0 0cm 0 .9cm},clip, width=\linewidth]{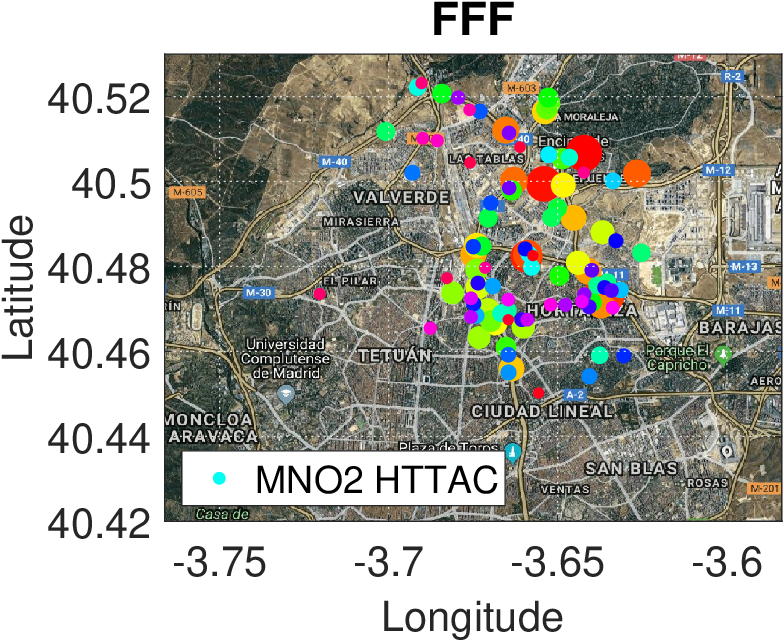}
		\caption{}
	\end{subfigure}
	\hfill
	\begin{subfigure}{0.45\textwidth} 
		\includegraphics[trim={0 0cm 0 0.9cm},clip, width=\linewidth]{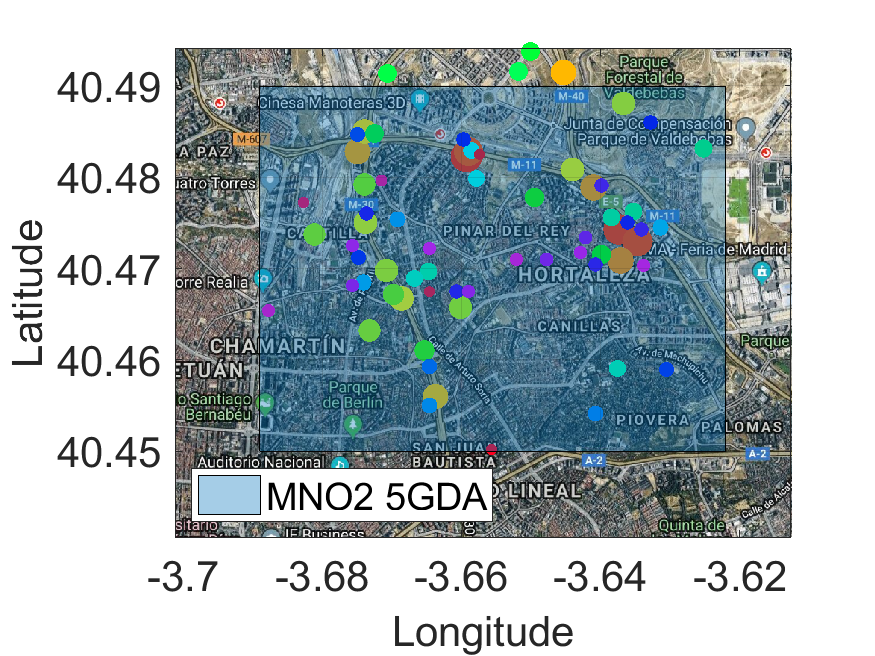}
		\caption{}
	\end{subfigure}
	
	\vspace*{-\baselineskip}
	\begin{subfigure}{0.45\textwidth}
		\includegraphics[trim={0 0cm 0 .9cm},clip, width=\linewidth]{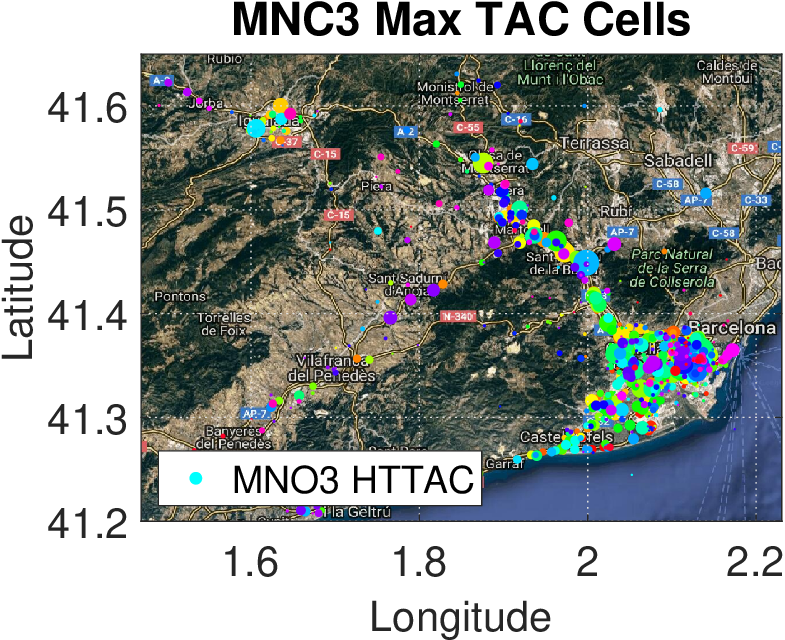}
		\caption{}
	\end{subfigure}
	\hfill
	\begin{subfigure}{0.45\textwidth} 
		\includegraphics[trim={0 0cm 0 0.9cm},clip, width=\linewidth]{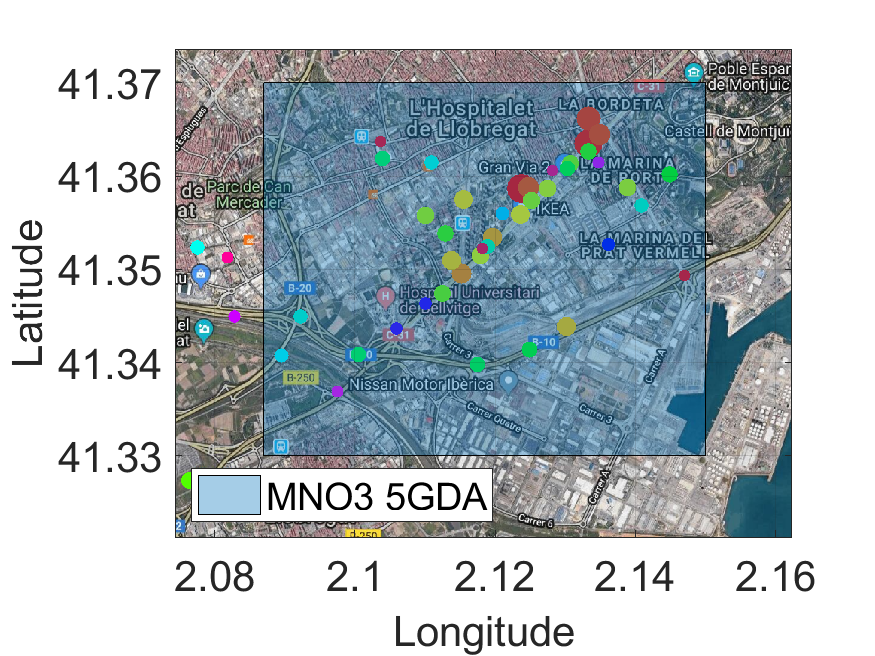}
		\caption{}
	\end{subfigure}
	
	\caption{Colored circles with largest radius represents the LTE cells with highest traffic (a) Highest traffic area represented by HTTACs (b) HTTAC of MNO1 showing only n=100 LTE cells (c) Demarcation of the 5GDA inside the highest traffic area} 
	\label{fig:HTTAC_5GDA}
\end{figure*}

Finally, the 5G deployment area selection ($A_m^g$) is determined. The network data analytics guide MNOs in the planning process to select those highly concentrated areas known as 5GDAs for new deployments. The HTTACs $H_1,H_2$ and $H_3$ for the three MNOs are shown in Fig.\ref{fig:HTTAC_5GDA}(a,b,c). We found that the areas covered by $H_1$ and $H_2$ both lie in the center of Madrid, while $H_3$ is located in the center of Barcelona (see Fig.\ref{fig:HTTAC_5GDA}(c)). To locate the 5GDA, we set $N_c=100$ to find the highest traffic cells residing within each $H_m$, as Fig.\ref{fig:HTTAC_5GDA}(b) shows for MNO1. The HTTAC of MNO1 exhibits a head/tail pattern, in which a high density of cells represents the head of the HTTAC and the tail represents the lower concentration of cells. The same pattern is observed for MNO3 in Fig.\ref{fig:HTTAC_5GDA} (c and f), where tail lies in the upper part of the HTTAC and head is found at the bottom. However, in case of MNO2 Fig.\ref{fig:HTTAC_5GDA} (b and e) the pattern is homogeneous, where cells are concentrated towards the center of the city and less along the road sides.

These type of patterns help the network planning engineer to locate the high-concentration sub-area $A_m^g  \subseteq  A_m^{N_c}$ for 5G deployments and perform the demarcation of the area as shown by rectangular boxes in Fig.\ref{fig:HTTAC_5GDA}(b,d,f) for three MNOs. The values of the manually demarcated areas are $A_m^g = \left[23.69,25.20,23.39\right]$ $\text{km}^{2}$ for MNO1, MNO2 and MNO3, respectively. The demarcated area $A_m^g$ is an essential input to the \textit{Network Dimensioning}. 


\section{CONCLUSIONS}
We show how network data extracted from public open repositories can assist in the network dimensioning phase. In particular, the OpenCellID database is considered because millions of samples corresponding to the legacy network of the MNOs are publicly available for research purpose. An interesting aspect of the selected database is how LTE identifiers can assist in manifold ways to obtain relevant information after processing raw data. Specifically, to form the database matrix $\mathbf{D}$, to understand the crowdsourced database and its features, to identify unique Cells and TAC IDs ($\mathcal{N}_m$ and $\mathcal{T}_m$) and to determine the highest traffic $N_c$ cells in the existing infrastructure per MNO. 

The proposed procedure infers information of the network infrastructure and traffic on the existing eNB cells that is utilized to determine the ID ($H_m$) of the HTTAC area. The traffic patterns in the determined HTTAC area guide the demarcation of the 5GDA before the dimensioning phase. For 5GDA, the financial constraint of the MNO is addressed through the value of $N_c$ cells, which restricts the number of cells to be included in the target area. As shown in the results, the geographical area $A_m^{N_c}$ of the $N_c$ cells reveals the traffic patterns in the target area. These traffic patterns then guide the network planning engineers for the appropriate and manual demarcation of the 5GDA ($A_m^g$) of the highest traffic density per MNO. As explained, learning about the existing network infrastructure through data can provide useful information about the traffic and corresponding areas of the highest density of subscribers. The output of the \textit{Network Data Acquisition} entity is the demarcated 5GDA ($A_m^g$) that serves as a primary input to the \textit{Network Dimensioning} entity.

Regarding open traffic databases, they offer many potential applications of the data they contain. However, they must be mapped and used carefully before a previous analysis, as these databases are populated through voluntary contributions and there is no guarantee of quality in the raw data they store. This situation is addressed with the Network Data Acquisition method described in this paper.

\addtolength{\textheight}{-12cm}

\section*{ACKNOWLEDGMENT}

\bibliographystyle{ieeetr}
\bibliography{Bibliography}

\end{document}